\documentclass[aps,amsmath,prl,twocolumn,a4paper,floatfix]{revtex4}
\usepackage{graphicx}
\usepackage{amsmath}
\usepackage{amssymb}
\usepackage{amsfonts}
\newcommand{\comment}[1]{}

\newcommand{\met}{\frac{1}{2}}

\begin{document}

\title{Spin Hall effect in a two-dimensional electron gas in the presence of a magnetic field}

\author{P. Lucignano$^{1}$, R. Raimondi$^2$, A. Tagliacozzo$^1$}
\address{$^1$  Coherentia-INFM and Dipartimento di Scienze Fisiche Universit\`a degli
             studi di Napoli "Federico II",Monte S.Angelo - via Cintia, I-80126 Napoli, Italy}
\address{$^2$CNISM and Dip. di Fisica "E. Amaldi", Via della Vasca Navale 84, 00146 Roma, Italy}
\date{\today}
\begin{abstract}
We study  the spin Hall effect  of a two-dimensional electron gas in the presence of a magnetic field and both the Rashba and Dresselhaus spin-orbit interactions. We show that the value of the spin Hall conductivity, which is finite  only if the Zeeman spin splitting is taken into account, may be tuned by varying the ratio of the in-plane and out-of-plane components of the applied magnetic field.  We identify the origin of this behavior with the different role played by the interplay of spin-orbit and Zeeman couplings for in-plane and out-of-plane magnetic field components. \end{abstract}
\maketitle
The spin Hall effect is the transverse pure spin current response to an applied electric field, with the spin current polarization perpendicular to both the spin current and the electric field. This effect in semiconductors is due to the spin-orbit coupling: for a review see Ref.\cite{engel2006}. In a disordered two-dimensional electron gas (2DEG), it is now accepted that, at least in uniform and stationary conditions, the   spin Hall conductivity (SHC) vanishes for the Rashba spin-orbit interaction or linear Dresselhaus spin-orbit coupling\cite{inoue2004,mishchenko2004,raimondi2005,khaetskii2006}. The reason for this has been clearly associated to the linear-in-momentum dependence of the spin-orbit interaction in a 2DEG\cite{rashba2004,dimitrova2005,chalaev2005}.
As pointed out by Rashba\cite{rashba2004}, disorder is only important in order to obtain a steady state and the vanishing of the SHC persists
in the presence of a perpendicular magnetic field if the Zeeman splitting is neglected.  On the other hand, in a disordered 2DEG, the presence of the Zeeman coupling due to an in-plane applied magnetic field, may lead to a non-vanishing SHC\cite{lin2006,engel2007,milletari2008}. Therefore, it is relevant to further investigate how the interplay of Zeeman coupling and various spin orbit interactions may lead to a finite SHC in a 2DEG in a magnetic field even in the absence of disorder. In this paper we show that the role played by the Zeeman coupling depends on whether the applied magnetic field has an in-plane component in addition to the one perpendicular to the plane of the 2DEG.
 
In the case of a perpendicular magnetic field,  Rashba provided an elegant proof of the vanishing of the SHC by deriving  a sum rule based on the knowledge of the exact eigenvalues and eigenstates of the Hamiltonian\cite{rashba2004}. When both the Rashba and Dresselhaus interactions are present an exact solution for the eigenstates and eigenvalues is only  known in terms of recurrence relations for the expansion  coefficients in the Landau states basis\cite{degang2006}. For the special case when the two spin-orbit interactions have equal strenghts one can again explicitly diagonalize the Hamiltonian\cite{zarea2005,wang2005,tarasenko2002,averkiev2004}.
In this case the SHC has been investigated in the presence of Zeeman coupling\cite{degang2008}.
However, when the Zeeman coupling is due to a perpendicular magnetic field, the SHC is associated to a charge Hall conductivity (CHC).  As  the Zeeman coupling gives rise to a perpendicular spin polarization,  it is not surprising that the CHC also carries a spin current.
In the case of a tilted magnetic field, we show that, in a clean sample, a new contribution to the SHC appears and, most interestingly, it is anisotropic with respect to the direction of the in-plane component.  This phenomenon is similar to what has been predicted for the case of  a disordered 2DEG.


 The Hamiltonian for the 2DEG reads
\begin{equation}
\label{eq1}
H=H_0+ H_{Z} + H_{SO},
\end{equation}
where the spin-independent part is given by
\begin{equation}
\label{eq2}
H_0=\frac{1}{2m} {\boldsymbol {\Pi}}^2, \ 
{\boldsymbol {\Pi}}=\mathbf{ p}+\frac{e}{c}\mathbf{ A}, \ e>0\;.
\end{equation}
Here the electrons are supposed to be confined in the $x-y$ plane.
It is useful to choose, for an arbitrarily oriented magnetic field $\mathbf{B}=(B_x,B_y,B_z)$, the  gauge $ \mathbf{ A} = (-B_z y+B_y z,-B_x z, 0)$. The Zeeman coupling is
\begin{equation}
\label{eqz}
H_Z= \met g \mu_B \mathbf{B} \cdot {\boldsymbol \sigma},
\end{equation}
where ${\boldsymbol \sigma}=(\sigma_x,\sigma_y,\sigma_z)$ is the Pauli matrix vector.
Finally, the spin-orbit Hamiltonian is given by
\begin{equation}
\label{eq3}
H_{SO}= \frac{\alpha}{\hbar} \left(\Pi_y \sigma_x - \Pi_x \sigma_y\right)+
\frac{\beta}{\hbar} \left(\Pi_y \sigma_y - \Pi_x \sigma_x\right),
\end{equation}
with the two  terms describing  the Rashba and the 
linear Dresselhaus spin-orbit interactions, respectively.\\

Due to our gauge choice, the system has translational invariance along the x-direction, $p_x=\hbar k$ is a good quantum number
and the orbital part of the eigenfunctions can be chosen as
\begin{equation}
\label{eq4}
\phi(x,y)=e^{ikx}\varphi(y-y_0),
\end{equation}
where $\varphi $ is a generic harmonic oscillator wavefunction and $y_0=l_b^2k$ indicates the orbit center coordinate with
$l_b=\sqrt{\hbar c /e B}$ the magnetic length.  By introducing the creation and annihilation operators
for the harmonic oscillator
\begin{eqnarray}
\Pi_x+ i \Pi_y&=&-  \sqrt{2}\hbar l_b a^\dagger\nonumber\\
\Pi_x- i \Pi_y&=&- \sqrt{2}\hbar l_b a,  \label{eq5}
\end{eqnarray}
and by measuring all the energies in units of the ciclotron energy
$\hbar \omega_c = \hbar eB_z /mc$, 
we get the Hamiltonian (\ref{eq1}) in the following dimensionless form
\begin{eqnarray}
H& = &  \left(a^\dagger a +1/2\right)+ \frac{g}{4}
\left(\sigma_z+ \frac{B_x}{B_z} \sigma_x+\frac{B_y}{B_z} \sigma_y\right) \label{eq6} \\
 & +& i  \sqrt{2} \eta_R
\left(a^\dagger\sigma^- -a
\sigma^+\right) + \sqrt{2}\eta_D \left(a^\dagger\sigma^+ +a
\sigma^-\right),\nonumber
\end{eqnarray}
where $\eta_R=\alpha m l_b/\hbar^2$ and $\eta_D=\beta m l_b/\hbar^2$. In order to analyze the role played by the various components of the magnetic field, it is useful to introduce 
 $g_x=g {B}_x/B_z$,  $g_y=g {B}_y/B_z$, and  $g_z=g$ 
to be treated as independent parameters. 

Following Rashba we compute the zero frequency  SHC, whose   Kubo formula reads
\begin{equation}
\label{eq7}
\Sigma_{zxy}=-\frac{ie}{\pi \hbar l_b^2} \lim_{\omega\to0}
\sum^*_{\lambda,\lambda'} 
\frac{
\Im m{ \left(
\langle \lambda|v_y |  \lambda' \rangle 
\langle \lambda'|J_{zx}|  \lambda\rangle\right)} }
{(\omega_{\lambda'}-\omega_{\lambda})^2-\omega^2},
\end{equation}
where $|\lambda\rangle$ are the eigenstates with eigenvalues $E_{\lambda}=\hbar \omega_{\lambda}$ and the star above  the sum indicates that we are only considering states $|\lambda'\rangle$ below the Fermi energy (here set to zero for simplicity) and $|\lambda\rangle$ above the Fermi energy.

The spin current operator,  for the z-component of the spin, is defined as
\begin{equation}
\label{eq8}
J_{zi}=\met(\sigma_z v_i+ v_i \sigma_z).
\end{equation}
In terms of the harmonic oscillator operators of Eq.(\ref{eq5}), the velocity operators can be derived according to $v_x =[x,H] /i\hbar$ and $v_y=[y,H]/i\hbar$. They read
\begin{eqnarray}
v_x&=& -\frac{\hbar}{m\sqrt 2 l_b} (a^\dagger+a+\sqrt 2
\eta_R \sigma_y + \sqrt 2 \eta_D \sigma_x)\label{eq9}\\
v_y&=& \frac{\hbar}{m\sqrt 2 l_b} (i(a^\dagger-a)+\sqrt 2
\eta_R \sigma_x + \sqrt 2 \eta_D \sigma_y) \nonumber\;,
\end{eqnarray}
so that  the spin current becomes
\begin{equation}
\label{eq11}
J_{zx}=-\frac{\hbar}{m \sqrt 2 l_b}
\left(a^\dagger+a\right)\sigma_z .
\end{equation}

Eq.(\ref{eq7}) together with Eq.(\ref{eq9}) is a generalization of Eq.(5)  of \cite{rashba2004}. 
In Ref.\cite{rashba2004} the author studies a case with only the Rashba spin-orbit interaction and a magnetic field orthogonal to the 2DEG in the absence of Zeeman splitting. In that case, the matrices $\langle \lambda|v_y |  \lambda' \rangle$ and  $\langle \lambda'|J_{zx}|  \lambda\rangle$ are antisymmetric and symmetric, respectively (with respect to the exchange of $\lambda$ and $\lambda'$), giving rise to a purely imaginary matrix element $\langle \lambda|v_y |  \lambda' \rangle \langle \lambda'|J_{zx}|  \lambda\rangle$.  Hence the two terms in the commutator appearing in the Kubo formula are equal and opposite  giving just a factor of two.
By contrast, we study the more general case in which also the Dresselhaus  and  Zeeman couplings are present. In this case no general argument about the symmetry of the matrix elements holds and, therefore, the imaginary part in Eq.(\ref{eq7}) has to be taken explicitly.

In order to numerically diagonalize the Hamiltonian we expand the eigenstates $|\lambda\rangle=\sum_{n\sigma} c^\lambda_{n\sigma} |n\sigma\rangle$in the Landau levels  (LL) basis
\begin{eqnarray}
|n,\sigma\rangle=\frac{(a^\dagger)^n}{\sqrt{n!}}|0\rangle \otimes|\sigma\rangle,
\end{eqnarray}
corresponding to the eigenvalues $E_n=\hbar \omega_c \left(n+1/2\right)$ in the absence of spin-orbit interactions and Zeeman couplings.
By working out the matrix elements, Eq.(\ref{eq7}) for the SHC becomes
\begin{equation}
\label{eq14}
\Sigma_{zxy}(0)=
-\met \frac{e\omega_c}{\pi \hbar } 
\sum^*_{\lambda,\lambda'}
\frac{1}{(\omega_{\lambda'}-\omega_{\lambda})}
\Re e{\left(A_{\lambda \lambda'}*B_{\lambda \lambda'}\right)},
\end{equation}
where
\begin{eqnarray}
A_{\lambda \lambda'}&=&\sum_{n\sigma} \sqrt{n} c^{\lambda *}_{n\sigma}c^{\lambda'}_{n-1\sigma} + 
\sqrt{n+1}c^{\lambda *}_{n\sigma}c^{\lambda'}_{n+1\sigma},\\
B_{\lambda \lambda'}&=&\sum_{n'} 
(\sqrt{n'+1} c^{\lambda' *}_{n'+1\uparrow}c^{\lambda}_{n'\uparrow}+
\sqrt{n'}   c^{\lambda' *}_{n'-1\uparrow}c^{\lambda}_{n'\uparrow}  \nonumber\\
&-& \sqrt{n'+1}   c^{\lambda' *}_{n'+1\downarrow}c^{\lambda}_{n'\downarrow}-
\sqrt{n'}   c^{\lambda' *}_{n'+1\downarrow}c^{\lambda}_{n'\downarrow} ).  
\end{eqnarray}

The numerical evaluation of Eq.(\ref{eq14})  shows that, in the absence of the Zeeman coupling, the SHC vanishes in all cases: whether the  Rashba, the Dresselhaus or both  spin-orbit couplings are present. This is a generalization of what obtained in Ref.\cite{rashba2004}. In that case it has been shown that the SHC is zero in the presence of Rashba spin-orbit coupling, with a sum rule argument.  That result  is also valid in the case of a Dresselhaus spin-orbit coupling, as this kind of interaction can be obtained, starting from the Rashba one, by use of the following rotation in the spin pace
\begin{eqnarray}
\label{symm0}
\sigma_x &\Rightarrow &\sigma_y   (\sigma^- \Rightarrow -i\sigma^+)\\
\sigma_y &\Rightarrow &\sigma_x   (\sigma^+ \Rightarrow i\sigma^-)\nonumber \\
\sigma_z& \Rightarrow &-\sigma_z \nonumber.
\end{eqnarray}
However, in spite of the fact that, when both interactions are present, we were not able to extend the  sum-rule arguments, our numerical calculation clearly shows that also in the presence of both kinds of spin orbit interactions the SHC vanishes. 

\begin{figure}[h]
\begin{center}
\includegraphics[width=\linewidth]{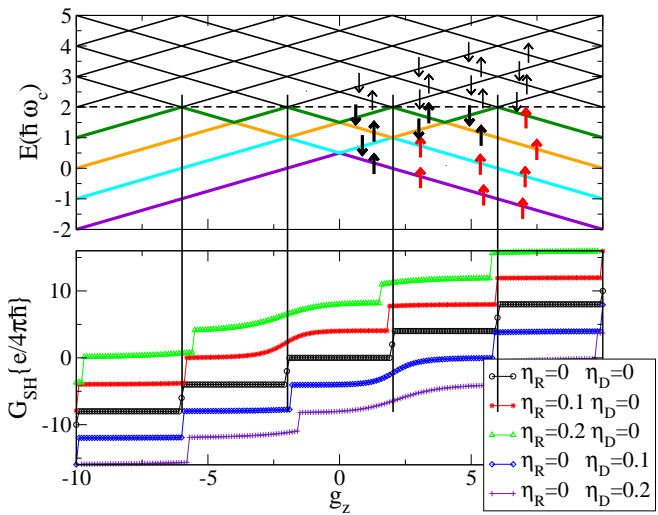}
\caption{(color on-line) (Upper panel) Low lying energy spectrum as function of the Zeeman coupling $g_z$. 
The arrows indicate the spin polarization of the LL's. Red arrows indicate the unbalanced spins below the Fermi levels. 
(Lower panel) SHC in units of $e/4\pi \hbar$ as function of $g_z$ for different values of $\eta_R$ and $\eta_D$ (indicated in the plot).  $\omega_c=2meV$}
\label{gradini}
\end{center}
\end{figure}

In order to study the effect of the Zeeman coupling on the SHC, we consider first the case with  the magnetic field orthogonal to the 2DEG
with no in-plane component.
In  Fig.\ref{gradini}  we plot the low lying part of the energy spectrum (upper panel) and the SHC (lower panel) as a function of $g_z$.
The energy spectrum shown corresponds to the case $\eta_R=\eta_D=0$. For simplicity, we have chosen, at  $g_z=0$, a Fermi level lying  between the second and the third LL, (the dashed line in the figure).
Here, and in the following, the cyclotron frequency is $\omega_c=2meV$ corresponding to a magnetic field of $\sim 1T$ for $GaAs$ based materials.  The corresponding SHC for $g_z=0$ is given by the black curve in the lower panel. It is quantized because the CHC  is, due to the choice of having an integer number of LLs fully occupied.  While the CHC is constant with $g_z$, the SHC shows sharp steps at those level crossings at which   the total spin polarization of the Fermi sea changes.
To emphasize how this happens  we have connected  in Fig.\ref{gradini} by vertical lines  the energy level crossings  with the steps of the SHC.
We can clearly see, by inspection, that only an odd number of crossings gives rise to a new net polarization of the current carrying states,  resulting in  the steps of the SHC of Fig.\ref{gradini}.  Of course, here spin and charge conductance quantizations go together. The other curves in the lower panel of Fig.\ref{gradini} describe the behavior of the SHC at finite values of $\eta_R$ \cite{nota} or $\eta_D$  and have been shifted uniformly for clarity.  A finite Rashba or Dresselhaus slightly modifies the shape of the  SHC steps, without substantially changing the scenario obtained in the absence of spin-orbit interaction.  In particular, there is a rounding effect and a small shift of the $g_z$ values at which the steps occur, which is positive or negative in the case of Dresselhaus and Rashba coupling, respectively.

\begin{figure}[!htb]
\begin{center}
\includegraphics[width=\linewidth]{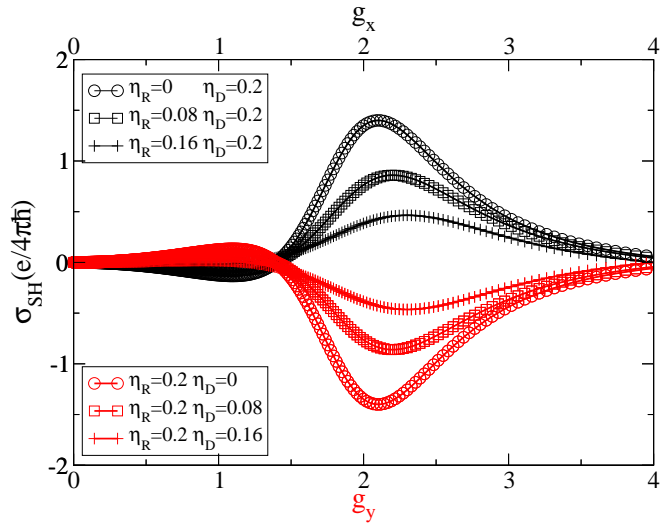}
\caption{(color on-line)Spin Hall conductivity in units of $e/4\pi \hbar$ as function of $g_Y$ or $g_X$ for different values of $\eta_R$ and $\eta_D$ (indicated in the plot). }
\label{gx-gy}
\end{center}
\end{figure}

\begin{figure}[!htb]
\begin{center}
\includegraphics[width=\linewidth]{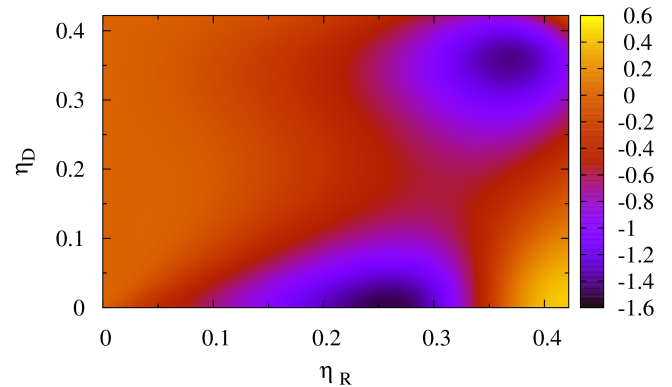}
\caption{(color on-line)SHC as a function of $\eta_R$ and $\eta_D$ for $g_x=0$, $g_y=2$}
\label{colorplot1}
\end{center}
\end{figure}

\begin{figure}[!htp]
\begin{center}
\includegraphics[width=\linewidth]{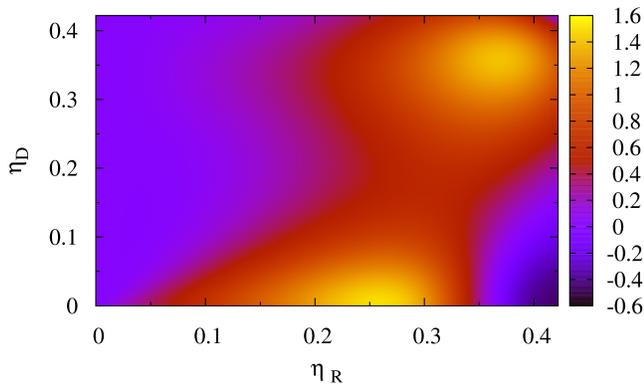}
\caption{(color on-line)SHC as a function of $\eta_R$ and $\eta_D$ for $g_x=2$, $g_y=0$}
\label{colorplot2}
\end{center}
\end{figure}

Next we turn to the case in which the magnetic field is tilted with respect to the z-axis and has both in-plane and out-of-plane components. Two different subcases are considered: a magnetic  field  lying  in the $xz$ plane or in the $yz$ plane.
Moreover, in order to emphasize the effect of the in-plane Zeeman coupling, we will first neglect  the effect of  $g_z$. Therefore the relevant Zeeman parameters are $g_x$ and $g_y$. These cases are analysed in detail in Figs. \ref{gx-gy}, \ref{colorplot1}, \ref{colorplot2}. 

In Fig.\ref{gx-gy} we plot the SHC at a fixed $\eta_R\neq0$ for  various $\eta_D$ (lower red curves) and at a fixed $\eta_D\neq0$ for various $\eta_R$ (top black curves). The former are plotted as a function of $g_y$ and the latter as a function of $g_x$. This is because 
 at {$\eta_R=0 ,\;\eta_D\neq0$} the SHC  is identically zero if $g_x=0$ for any $g_y\neq0$, while at {$\eta_R\neq0 ,\;\eta_D=0$}, the SHC  vanishes for any  $g_x\neq 0$ if $g_y=0$.
The vanishing of the SHC at $g_y=g_x=0$ has been found in ref.\cite{rashba2004}.
At increasing $g_{y(x)}$ the peak in the SHC is larger the larger is the difference  between $\eta_R$ and $\eta_D$. 


The curves in Fig.\ref{gx-gy} show a remarkable reflection symmetry with respect to the x-axis. Indeed,   the Rashba Hamiltonian, in the presence of a $g_y$ Zeeman splitting, maps into the Dresselhaus Hamiltonian in the presence of a $g_x$ Zeeman splitting.  However this property is even more general, provided that $g_z$ vanishes, as the unitary transformation of Eq.(\ref{symm0}) maps the Hamiltonian into itself by interchanging the pair $(\eta_R,g_y)$ with the pair $(\eta_D,g_x)$.
While the velocity operator $\mathbf{ v}$ does not change under this transformation,
the current $J_{zy}$ changes its sign. As a result the SHC is antisymmetric with respect to the interchange of $\eta_R,g_y$ with $\eta_D,g_x$ and therefore
\begin{eqnarray}
\label{symmetry2}
&\Sigma_{zxy}&(\eta_R=a,\eta_D=b,g_x=c,g_y=d)=\\
=-&\Sigma_{zxy}&(\eta_R=b,\eta_D=a,g_x=d,g_y=c).\nonumber
\end{eqnarray}
This result has already been found in the absence of magnetic field \cite{shen2004}.
In Figs. \ref{colorplot1}, \ref{colorplot2}, the color maps refer to the SHC for a wide window of  $\eta_R$,  $\eta_D$ (x and y axis).  In Fig \ref{colorplot1} $g_x=0$ and $g_y=2$, while Fig.\ref{colorplot2} shows the complementary case   $g_x=2$ and $g_y=0$. 
The color pattern in the figures \ref{colorplot1} and \ref{colorplot2} is exactly the same but with the interchange of the sign of the SHC because of the symmetry property (\ref{symmetry2}).
It can be noticed  that  the SHC  can be sizeably different from zero in a wide range of parameters. 
We stress that the SHC shown here is not associated with a spin polarized charge Hall current as, up to now, we have been considering just the case  $g_z=0$.
As $g_x$ and $g_y$ include the components of the magnetic field in their definition, exchanging $g_x$ and $g_y$ is equivalent to a change of direction of the in-plane magnetic field. This implies that  when the magnetic field is in the $x$ direction a nonzero SHC only occurs for finite $\eta_D$, while when the field is in the $y$ direction the SHC occurs for finite $\eta_R$ only.
\begin{figure}[!htp]
\begin{center}
\includegraphics[width=\linewidth]{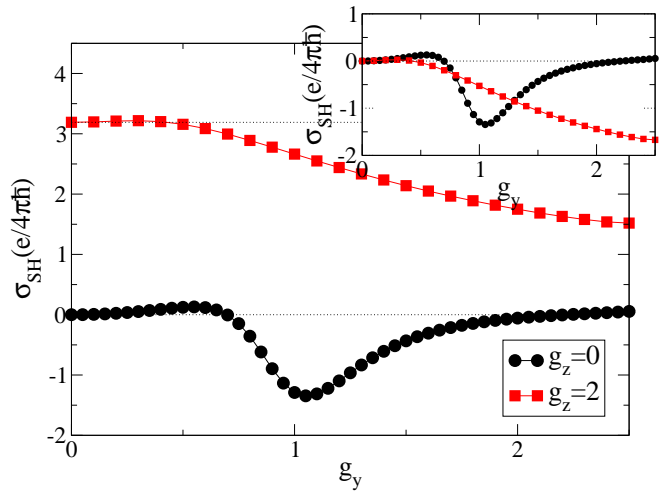}
\caption{(color on-line) Non additivity of the SHC with $g_y$ and $g_z$}
\label{noadd}
\end{center}
\end{figure}

However, in a real experiment of a 2DEG with both perpendicular and parallel magnetic fields, one has also the Zeeman spin splitting in the z-direction parametrized by a non-vanishing $g_z$. We conclude, then, our analysis, by  discussing this case. We choose the magnetic field oriented in the $y-z$ plane, as the $x-z$ orientation can be inferred by symmetry, according to Eq.(\ref{symmetry2}).
The reference curve, for $g_z=0$ is plotted in the lower part od Fig.\ref{noadd}. The SHC is non vanishing only in a restricted region of $g_y$ values because, in the absence of $g_z$, the electron spins tend to polarize along the $y$-axis,  when $g_y$ grows, so
inhibiting the Rashba spin precession. As a result the SHC tends to zero since no $z$-axis polarization is available for the spin current.
This does not happen when $g_z$ is finite as it appears in the upper curve of Fig.\ref{noadd}: there, the competition of the two orthogonal Zeeman splittings ($g_y$ and $g_z$) does not allow for  an in-plane polarization effect, therefore letting the Rashba (or Dresselhaus) coupling playing its role.
The measured SHC, in this case, will have two contributions: the first one is due the Zeeman polarization of the LL discussed so far in Fig. \ref{gradini}, the second (the most interesting one) is the SHE due to the spin-orbit coupling of Figs.\ref{gx-gy},\ref{colorplot1},\ref{colorplot2}. 
As it can be seen from the Fig. \ref{noadd}, the two effects do not superimpose trivially, {\emph i.e.},  the SHC in the presence of both a Zeeman interaction in the $z$ and  $y$ direction is not the linear superposition of the SHC with $g_y$ or $g_z$ only.   To stress this point we have  plotted  the SHC as a function of $g_y$ in Fig. \ref{noadd}, with $g_z=0$ (black curve)  and $g_z=2$ (red curve) for $\eta_R=0.2$ (and $\eta_D=0$ for simplicity).
The behavior of the two curves as function of $g_y$ is substantially different. The striking result is that a finite $g_z$ makes  the SHC finite even at increasing values of $g_y$, in a region where the SHC would vanish if $g_z=0$. This is clearly seen in the inset of Fig. \ref{noadd}, where the red curve has been shifted downward to eliminate the \textit{background} SHC associated with the cohexisting CHC with a finite $g_z$.  Of course,  the delocalization of the
orientation of the spin, due to  presence of both non commuting spin terms in the Hamiltonian, makes the tilting of the spin in the z-direction
easier even at larger $g_y$.

In a real sample, one expects that both $g_y(x)$ and $g_z$ are different from zero and in a nonzero SHC can be expected when  a tilted magnetic field is applied.

In conclusion, we have shown that a finite spin Hall effect is possible when both out-of-plane and in-plane magnetic field components are applied to a 2DEG in the presence of  Rashba and/or Dresselhaus spin-orbit couplings. The interplay of the Zeeman couplings along the perpendicular and parallel direction, with respect to the plane of the 2DEG, allows for a tuning of the SHC and should provide the way for an experimental detection. 

Depending on whether the Rashba or Dresselhaus spin orbit coupling dominates, the SHC changes with changing of  the orientation of the in-plane component of the magnetic field.

We acknowledge useful discussions with Vincenzo Marigliano Ramaglia and financial support from CNISM under Progetti Innesco 2006  as well as from CNR-INFM within ESF Eurocores Programme FoNE (Contract No. ERAS-CT-2003-  980409).


\begin{references}
\bibitem{nota} We fixe $\omega_c=2meV$ therefore  also $l_b$ is fixed and $\eta_R=0.2$ for $\alpha=100meV\AA$
\end{references}
\end{document}